\documentclass[onecolumn,12pt]{article}
\usepackage{graphicx}
\begin{document}
\baselineskip 14pt

\title{Distortion of Magnetic Fields in a Starless Core II: \\
3D Magnetic Field Structure of FeSt 1-457} 
\date{}
\author{Ryo Kandori$^{1}$, Motohide Tamura$^{1,2,3}$, Kohji Tomisaka$^{2}$, Yasushi Nakajima$^{4}$\\
Nobuhiko Kusakabe$^{3}$, Jungmi Kwon$^{5}$, Takahiro Nagayama$^{6}$, Tetsuya Nagata$^{7}$, \\
and \\
Ken'ichi Tatematsu$^{2}$\\
{\small 1. Department of Astronomy, The University of Tokyo, 7-3-1, Hongo, Bunkyo-ku, Tokyo, 113-0033, Japan}\\
{\small 2. National Astronomical Observatory of Japan, 2-21-1 Osawa, Mitaka, Tokyo 181-8588, Japan}\\
{\small 3. Astrobiology Center of NINS, 2-21-1, Osawa, Mitaka, Tokyo 181-8588, Japan}\\
{\small 4. Hitotsubashi University, 2-1 Naka, Kunitachi, Tokyo 186-8601, Japan}\\
{\small 5. Institute of Space and Astronautical Science, Japan Aerospace Exploration Agency,}\\
{\small 3-1-1 Yoshinodai, Chuo-ku, Sagamihara, Kanagawa 252-5210, Japan}\\
{\small 6. Kagoshima University, 1-21-35 Korimoto, Kagoshima 890-0065, Japan}\\
{\small 7. Kyoto University, Kitashirakawa-Oiwake-cho, Sakyo-ku, Kyoto 606-8502, Japan}\\
{\small e-mail: r.kandori@nao.ac.jp}}
\maketitle

\abstract{
Three dimensional (3D) magnetic field information on molecular clouds and cores is important for revealing their kinematical stability (magnetic support) against gravity which is fundamental for studying the initial conditions of star formation. In the present study, the 3D magnetic field structure of the dense starless core FeSt 1-457 is determined based on the near-infrared polarimetric observations of the dichroic polarization of background stars and simple 3D modeling. With an obtained angle of line-of-sight magnetic inclination axis $\theta_{\rm inc}$ of $45^{\circ}\pm10^{\circ}$ and previously determined plane-of-sky magnetic field strength $B_{\rm pol}$ of $23.8\pm12.1$ $\mu{\rm G}$, the total magnetic field strength for FeSt 1-457 is derived to be $33.7\pm18.0$ $\mu{\rm G}$. The critical mass of FeSt 1-457, evaluated using both magnetic and thermal/turbulent support is ${M}_{\rm cr} = 3.70\pm0.92$ ${\rm M}_{\odot}$, which is identical to the observed core mass, $M_{\rm core}=3.55\pm0.75$ ${\rm M}_{\odot}$. We thus conclude that the stability of FeSt 1-457 is in a condition close to the critical state. Without infalling gas motion and no associated young stars, the core is regarded to be in the earliest stage of star formation, i.e., the stage just before the onset of dynamical collapse following the attainment of a supercritical condition. These properties would make FeSt 1-457 one of the best starless cores for future studies of the initial conditions of star formation. 
}


\vspace*{0.3 cm}

\section{Introduction}
Three dimensional (3D) information on astronomical objects is essential for discussing their morphology, diversity, origin, and evolution, as well as for determining their physical parameters. For dense molecular clouds and their cores, it is important to know the inclination angle of the magnetic field direction (magnetic axis) toward the line of sight to investigate their magnetic support against self-gravity through the measurements of the mass-to-magnetic flux ratio compared with the critical value suggested by theory (e.g., Mestel \& Spitzer 1956; Nakano \& Nakamura 1978). 
We have presented observational results for the dense starless molecular cloud core FeSt 1-457 using the plane-of-sky magnetic field components in a previous paper (Kandori et al. 2017, hereafter Paper I). The value of the mass-to-magnetic flux ratio determines the stability of the core, and if the total magnetic field strength can be accurately determined using 3D magnetic field information, i.e., line-of-sight field inclination angle, the stability as well as evolutionary status of the core can be discussed. 
\par
To date however, the line-of-sight inclination angle of the magnetic field has not been extensively measured, and the ratio of the observational mass-to-flux ratios to the theoretical critical value, ${\lambda} = ({M}/{\Phi})_{\rm obs} / ({M}/{\Phi})_{\rm critical}$, have been determined mostly based on statistical considerations (e.g., Crutcher 2004; Troland \& Crutcher 2008). Due to difficulties in observation and/or accurate modeling, the number of molecular clouds or dense cloud cores with known 3D magnetic field structures is limited (e.g., Goodman \& Heiles 1994; Gon\c{c}alves, Galli, \& Girart 2008). 
\par 
The most straightforward way to measure the line-of-sight magnetic inclination axis is to combine measurements of the line-of-sight (e.g., Zeeman splitting observations of molecular species) and plane-of-sky (e.g., Chandrasekhar--Fermi method: Chandrasekhar \& Fermi 1953) magnetic field components. However, there are difficulties in combining magnetic field measurements obtained by different methods, because each methods has different spatial resolutions, probing depths, necessary measurement conditions, and other characteristics. A simple combination may reduce accuracy in determining the total magnetic field strength toward the clouds. Therefore, independent measurements of the magnetic field inclination angle are necessary, although there is a study of careful estimations of the 3D magnetic field structure of dark clouds using both Zeeman results and the Chandrasekhar--Fermi results (e.g., Goodman \& Heiles 1994). \par
There are limited methods for 3D magnetic field determination generally applicable for molecular clouds and cores. One is the model introduced by Myers \& Goodman (1991), which can estimate a 3D uniform magnetic field and its line-of-sight inclination by combining maps of both the plane-of-sky polarization angle distribution and the line-of-sight magnetic field component based on Zeeman observations. 
\par
For magnetic fields with a symmetric structure, a direct comparison with theoretical models can be worthwhile. The hourglass-shaped magnetic field around the protostar NGC 1333 IRAS 4A (Girart, Rao, \& Marrone 2006) was compared with theoretical models of the collapse of the magnetized cloud (Gon\c{c}alves, Galli, \& Girart 2008), resulting in a determination of the 3D magnetic structure around the protostar as well as the line-of-sight magnetic inclination angle $\theta_{\rm inc}$ of $0$--$30$ deg or $25$--$55$ deg. The relatively large difference in the obtained inclination angles is due to differences in the employed models. \par
In the present study, 3D magnetic fields around the nearby dense starless core FeSt 1-457 were determined through modeling using a simple 3D function. The only assumption is axial symmetry of the magnetic field in the core. The line-of-sight magnetic inclination angle could be successfully estimated with relatively small uncertainty. 
\par
FeSt 1-457 was observed with near-infrared (NIR) polarimetry, and a distinct symmetric hourglass-shaped magnetic field geometry associated with the core was revealed in 2D (Figure 1, taken from Paper I). The fundamental physical parameters of FeSt 1-457 were determined based on the density structure modeling using the Bonnor--Ebert model (Bonnor 1956; Ebert 1955). The radius, mass, and central density of the core are $18500\pm1460$ AU (144$''$), $3.55\pm0.75$ $M_{\rm \odot}$, and $3.5(\pm0.99) \times 10^{5}$ ${\rm cm}^{-3}$ (Kandori et al. 2005), respectively, at a distance of $130^{+24}_{-58}$ pc (Lombardi et al. 2006). 
\par
The plane-of-sky magnetic field strength, $B_{\rm pos}$, is $23.8\pm12.1$ $\mu{\rm G}$, which makes the core a magnetically supercritical state with $\lambda = 2.00\pm0.42$ (Paper I). The stability of the core can be evaluated by comparing the theoretical critical mass considering magnetic and thermal/turbulent contributions in the core of $M_{\rm cr} \simeq M_{\rm mag}+M_{\rm BE}$ (Mouschovias \& Spitzer 1976; Tomisaka, Ikeuchi \& Nakamura 1988; McKee 1989), with the observed mass of the core, $M_{\rm core}$, where $M_{\rm mag}$ is the magnetic critical mass and $M_{\rm BE}$ is the Bonnor--Ebert mass. The critical mass of the core is $M_{\rm cr} = 2.96\pm0.62$ $M_{\rm \odot}$, which is less than or comparable to the actual mass ($3.55\pm0.75$ $M_{\rm \odot}$) of the core, suggesting that the core is in a nearly critical state. 
\par
The magnetic field studies presented in Paper I are based on plane-of-sky magnetic fields, providing a lower limit of the total magnetic field strength. Therefore, accurate estimations of the magnetic inclination angle of the core as well as the whole magnetic field structure in 3D are necessary for further discussion of the magnetic field structure of FeSt 1-457. 

\section{Data and Methods}
The NIR polarimetric data for the 3D magnetic field analysis of FeSt 1-457 is taken from Paper I. The observations were conducted using the $JHK$${}_{\rm s}$-simultaneous imaging camera SIRIUS (Nagayama et al. 2003) and its polarimetry mode SIRPOL (Kandori et al. 2006) on the IRSF 1.4 m telescope at the South African Astronomical Observatory (SAAO). SIRPOL can provide deep- (18.6 mag in the $H$ band, $5\sigma $ in one hour exposure) and wide- ($7.\hspace{-3pt}'7 \times 7.\hspace{-3pt}'7$ with a scale of 0$.\hspace{-3pt}''$45 ${\rm pixel}^{-1}$) field NIR polarimetric data. 
\par
The polarimetry data toward the core consists of a superposition of the polarizations from the core itself and the ambient medium which is unrelated to the core. After subtracting the ambient polarization components, 185 stars located within the core radius ($R \le 144''$) in the $H$ band were selected for the 3D polarization analysis (Figure 1). \par
A 3D version of the simple function employed in Paper I, $z(r, \varphi, g) = g + gC{r}^{2}$ in cylindrical coordinates $(r, z, \varphi)$, was used to model the core magnetic fields, where $g$ specifies the magnetic field line, $C$ is the curvature of the lines, and $\varphi$ is the azimuth angle (measured in the plane perpendicular to $r$). In the function, the shape of the magnetic field lines is axially symmetric around the $r$ axis. The function $z(r, \varphi, g)$ has thus no dependence on the parameter $\varphi$. 
\par
First, the 3D unit vectors of the polarizations following the function were calculated with $500^3$ cells. These unit polarization vectors were then scaled to obtain the polarization vectors in each cell, $\Delta P_{\rm model}$. To determine the scaling factor in each cell, it is necessary to prepare the volume density values in each cell and the density--polarization conversion factor. The volume density in each cell can be obtained from known density structure of FeSt 1-457. The core's density structure was reported to be well fitted using the Bonnor--Ebert model with a best-fit Bonnor--Ebert solution parameter $\xi_{\rm max} = (R/C_{\rm s}) \sqrt{4 \pi G \rho_{\rm c}}= 12.6$, where $R$ is the core radius, $C_{\rm s}$ is the effective sound speed, $G$ is the gravitational constant, and $\rho_{\rm c}$ is the central volume density (Kandori et al. 2005). The density--polarization conversion factor was estimated based on the slope on the $P_H$ vs. $H-K_s$ diagram of $4.8$ $\%$ ${\rm mag}^{-1}$ (Paper I). The relationship of $P_H = 0.22 \times N({\rm H}_2)/(9.4 \times 10^{20})$ was obtained, where $N({\rm H}_2)$ is the volume density of molecular hydrogen. For converting the $H-K_s$ color excess, $E_{H-K_s}$, into $N({\rm H}_2)$, the relationships of $A_V = 21.7 \times E_{H-K_{s}}$ (Nishiyama et al. 2008) and $N({\rm H}_2)/A_V = 9.4 \times 10^{20}$ cm$^{-2}$ mag$^{-1}$ (Bohlin, Savage, \& Drake 1978) were used. 
\par
%
The polarization vector $\Delta P_{\rm model}$ was then converted into $\Delta Q/I$ and $\Delta U/I$, the Stokes $Q/I$ and $U/I$ parameters in each cell. The known inclination angle of 179$^{\circ}$ (measured from north of east on the plane of the sky, which is on the $r-z$ plane) obtained in Paper I was applied, and the inclination toward the line-of-sight direction ($\theta_{\rm inc}$) was treated as a free parameter. For sampling, $30^3$ cells were used, and after rotation of both plane-of-sky and line-of-sight direction, line-of-sight integrations of the cubes of $\Delta Q/I$ and $\Delta U/I$ were conducted to obtain model polarization vector maps projected on to the sky plane. The model maps obtained at each inclination angle can be compared with the observations. 
\section{Results and Discussion}
\subsection{3D Parabolic Model}
Figure 2 shows the polarization vector maps of the 3D parabolic model. The white line shows the polarization vector, and the background color and color bar show the polarization degree of the model core. In constructing the model core, the assumed density structure is the same as that for the Bonnor--Ebert solution fitted for FeSt 1-457 ($\xi_{\rm max}=12.6$, Kandori et al. 2005), and the 3D magnetic curvature is set to $C=2.5\times10^{-4}$ ${\rm arcsec}^{-2}$ for display. The applied line-of-sight inclination angle is labeled in the upper-left corner of each panel. \par
As shown in Figure 2, the polarization vector maps change dramatically depending on the inclination angle. There are four major characteristics. First, the maximum polarization degree in each map decreases from a $\theta_{\rm inc}=90^{\circ}$ (plane-of-sky magnetic axis, edge-on) to a $\theta_{\rm inc}=0^{\circ}$ (line-of-sight magnetic axis, pole-on) geometry. Second, the polarization vector map of $\theta_{\rm inc}=90^{\circ}$ geometry is very similar to the 2D parabolic shape, whereas, the map of $\theta_{\rm inc}=0^{\circ}$ geometry clearly shows a radial polarization vector pattern. Third, depolarization occurs at the equatorial plane, which is apparent in the polarization degree distribution in the panel for $\theta_{\rm inc}=30^{\circ}$ or $\theta_{\rm inc}=15^{\circ}$. The depolarization region can be seen as dark patches where the polarization degree is low compared with neighboring regions. Since we virtually observe the model core with distorted magnetic fields from an inclined direction, crossing of the polarization vector direction occurs for the vectors located at the front and back sides of the core. Note that this effect occurs over the whole equatorial plane, although the \lq \lq shadow'' region is prominent at a large distance from the center of the core. Finally, the shape of the polarization degree distribution is axially symmetric at $\theta_{\rm inc}=90^{\circ}$, but it gradually gets elongated in the north--south direction for decreasing $\theta_{\rm inc}$. This is also due to the depolarization effect in the core. \par
These characteristics of the 3D parabolic model make it possible to determine which inclination angle matches the observations. 
Similarly, Tomisaka (2011) and Kataoka, Machida, \& Tomisaka (2012) have shown that polarization images of the physical simulations of star-forming core projected on to the sky strongly depend on the viewing angle of the observations. 
Since the panels in Figure 2 differ from each other in both polarization degree and angle distributions, $\chi^{2}$ fitting with the observational data can be used to restrict the line-of-sight magnetic inclination angle as well as 3D magnetic curvature. 
\subsection{$\chi^{2}$ fitting}
Figure 3 shows the distribution of $\chi^2_\theta = (\sum_{i=1}^n (\theta_{\rm obs,{\it i}} - \theta_{\rm model}(x_i,y_i))^2 / \delta \theta_i^2$, where $n$ is the number of stars ($n=185$), $x$ and $y$ are the coordinates of the $i$th star, $\theta_{\rm obs}$ and $\theta_{\rm model}$ denote the polarization angle from observations and the model, and $\delta \theta_i$ is the observational error) calculated using the model and observed polarization angles. The best magnetic curvature parameter, $C$, was determined at each inclination angle $\theta_{\rm inc}$, and $\chi^2_\theta$ was obtained. 
From this polarization angle fitting, it is clear that the inclination angle is unlikely to have a $\theta_{\rm inc}$ value of $0^{\circ}$--$30^{\circ}$ because the angle of the model vectors shows a severe radial pattern (see Figure 2), which is not the case for the observed vectors (Figure 1). On the one hand, the high $\theta_{\rm inc}$ region of $>30^{\circ}$ cannot be excluded in the same way, because the radial pattern disappears and the model polarization angles converge to a parabolic shape similar to the 2D parabolic distribution. \par
We thus made further $\chi^2$ calculations using both the polarization angles and degrees. We determined the best model parameters at each $\theta_{\rm inc}$ by minimizing the difference in the polarization angles. We then calculated $\chi^2_P = (\sum_{i=1}^n (P_{\rm obs,{\it i}} - P_{\rm model}(x_i,y_i))^2 / \delta P_i^2$, where $P_{\rm obs}$ and $P_{\rm model}$ represent the polarization degree from observations and the model, and $\delta P_i$ is the observational error) in polarization degrees (Figure 4). In the procedure, the relationship between the model core column density and the polarization degree was scaled to be consistent with the observations. 
\par
In Figure 4, the minimization point is $\theta_{\rm inc} = 45^{\circ}$, and $\chi^2_P$ shows large values toward higher and lower inclination angles. The high $\theta_{\rm inc}$ region can be rejected, because of the lack of a depolarization region and the elongation of the polarization degree distribution. The number and strength of polarization vectors are greater in the North-South part of the core than those in the East-West part (Figure 1). 
Note that the depolarization effect occurs over the whole equatorial plane, and strong polarization vectors near the central region also suffer the depolarization effect. 
The low $\theta_{\rm inc}$ region can be rejected because of the shape of the depolarization region, the degree of the elongated shape in the polarization distribution, and the radial polarization distribution. 
\par
The 1-sigma error estimated at the minimum $\chi^2_P$ point is $6^{\circ}$, but based on the shape of $\chi^2_P$ distribution, we conclude that the magnetic inclination angle of FeSt 1-457 is $\theta_{\rm inc} = 45^{\circ} \pm 10^{\circ}$. The magnetic curvature obtained at $\theta_{\rm inc} = 45^{\circ}$ is $C = 2.0 \times 10^{-4}$ ${\rm arcsec}^{-2}$, four times steeper than that in the case of the 2D fitting (Paper I). 
The results of $\chi^2$ analysis presented here depend on the pattern of depolarization in the core. When the distributions of polarization angle are identical between observations and models, the difference of depolarization pattern is crucial in order to discriminate the solutions as shown in the two types of $\chi^2$ analysis in this section. Though these analysis is equivalent to calculate $\chi^2$ in $Q/I$ and $U/I$, the physical meaning is clear in the current $\chi^2$ analysis based on $P$ and $\theta$. 
\par
It is possible that the depolarization pattern generated by the inclined distorted field is affected by the phenomenon which is unrelated to the hourglass magnetic fields. 
As shown in Figure 8 of Paper I, FeSt 1-457 shows elongated column density distribution especially in the central part of the core. This does not affect the results of $\chi^2$ analysis, because the spatial extent of elongated central part of the core is small ($\approx 30''$) and no stars was detected in polarization in the region. 
If the grain alignment efficiency systematically varies in the core, this can affect the depolarization pattern analysis based on $\chi^2$. This is not the case for FeSt 1-457, because there is linear relationship between $P_H$ and $H-K_s$ as shown in Figure 6 of Paper I. 
The lack of uniformity of integrated materials toward far background stars can be the source of systematic polarization patterns which mimic depolarization. This is also not the case for FeSt 1-457. The radio observations of the core (Kandori et al. 2005) showed that there is only a single gas component toward the core. Furthermore, the polarization vectors surrounding the core are fairly uniform, and the subtraction analysis of the \lq\lq off-core'' component is successful as shown in Paper I. 
%
There is no specific polarization feature in and around the core except for the polarizations solely associated with the core. Therefore, the depolarization pattern observed toward FeSt 1-457 is most likely due to the effect of the crossing of polarization vectors located at the front and back side of inclined distorted fields surrounding the core. 
\par
Note, however, that if there is radiation anisotropy in the core, an angle-dependent radiative grain alignment may be taking place, and this may affect the polarization vectors toward the core. Such phenomenon is reported in the diffuse interstellar medium around a bright star (e.g., Andersson et al. 2011; Andersson \& Potter 2010), and a similar mechanism may work in the environment of a lack of radiation produced by the starless core. The linearity in $P_H$ vs. $H-K_s$ relationship cannot reject the possibility of the mechanism, because the radiative torque effect is an angular one. This effect could potentially create an angular polarization pattern around the core that would mimic the depolarization effect assumed in this work.  
\par
%
%
Note that the $\chi^2$ values in Figures 3 and 4 are relatively large. This seems to be due to the existence of polarization angle scatter mainly caused by Alfven wave, which can not be included in the observational error term in the calculation of $\chi^2$. 
Some contributions can come from the imperfect uniformity of integrated material toward the background stars, although the polarization component around the core is uniform enough to be subtracted off using a simple linear fitting. 
The main conclusion does not change if we replace the error term with a constant value representing the polarization angle scatter by Alfven wave. \par
Figure 5 shows the best-fit 3D parabolic model with the observed polarization vectors. The direction of the model polarization vectors is in good agreement with the observations. 
The standard deviation of the angular difference of the plane-of-sky polarization angles between the 3D model and the observations is $10.41^{\circ}$, which is consistent with the case of 2D model fitting (Paper I). 
There is a slight difference between the 3D model and observations in polarization degrees, particularly in the south-west region of the core, which may be due to an incomplete assumption for the axially symmetric magnetic fields. 

\subsection{Magnetic Properties of the Core}
For an obtained magnetic inclination angle $\theta_{\rm inc}$ of $45^{\circ}\pm10^{\circ}$, the total magnetic field strength of FeSt 1-457 is determined to be $B_{\rm pos}/\sin (\theta_{\rm inc})=23.8/\sin(45^{\circ})=33.7\pm18.0$ $\mu{\rm G}$. Note that the value is the magnetic field strength averaged for the whole core.
The ability of the magnetic field to support the core against gravity is investigated using the parameter ${\lambda} = ({M}/{\Phi})_{\rm obs} / ({M}/{\Phi})_{\rm critical}$, which represents the ratio of the observed mass-to-magnetic flux to a critical value $(2\pi {G}^{1/2})^{-1}$, suggested by theory (Mestel \& Spitzer 1956; Nakano \& Nakamura 1978). 
We find $\lambda = 1.41\pm0.38$, and the magnetic critical mass of the core is $M_{\rm mag}=1.77/\sin(45^{\circ})=2.51\pm0.86$ ${\rm M}_{\odot}$, which is still lower than the observed core mass of $M_{\rm core}=3.55\pm0.75$ ${\rm M}_{\odot}$. FeSt 1-457 is thus determined to be magnetically supercritical. 
The critical mass of FeSt 1-457, evaluated using both magnetic and thermal/turbulent support, $M_{\rm cr} \simeq M_{\rm mag}+M_{\rm BE}$ (Mouschovias \& Spitzer 1976; Tomisaka, Ikeuchi \& Nakamura 1988; McKee 1989), is $2.51+1.19=3.70\pm0.92$ ${\rm M}_{\odot}$, where $1.19\pm0.32$ ${\rm M}_{\odot}$ is the Bonnor--Ebert mass calculated using the kinematic temperature of the core of 9.5 K (Rathborne et al. 2008), the turbulent velocity dispersion of $0.0573\pm0.006$ ${\rm km}$ ${\rm s}^{-1}$, and the external pressure of $1.1(\pm0.3) \times 10^5$ ${\rm K}$ ${\rm cm}^{-3}$ (Kandori et al. 2005). We thus conclude that the stability of FeSt 1-457 is in a condition close to the critical state, with $M_{\rm cr} \approx M_{\rm core}$. 
\par
It is now known that FeSt 1-457 is a critical core. The critical nature of the core is consistent with (1) the starless characteristic of the core (Forbrich et al. 2009), (2) a lack of detection of supersonic infalling gas motion (Aguti et al. 2007), and (3) the detection of possible oscillating gas motion in the outer layer of the core (Aguti et al. 2007). Further magnetic diffusion and/or turbulent dissipation can eventually trigger the collapse of the core (e.g., Shu 1977; Nakano 1998). Note that external compression of the core (e.g., Frau et al. 2015 for the possibility of the cloud-cloud collision in the Pipe Nebula) may also initiate disrupt the star formation in the core. 
\par
The core may be kinematically sustained by some states close to a magneto-hydrostatic equilibrium structure. FeSt 1-457 shows an elongated structure in the central region on the dust extinction map (Figure 8 of Paper I or Figure 2 of Kandori et al. 2005), with the direction of the elongation perpendicular to the magnetic axis of the core. This is consistent with the theoretical perspective; a magneto-hydrostatic configuration (in a case with $M_{\rm mag}$ larger than $M_{\rm BE}$) can produce a flattened density distribution in the central region whose major axis is perpendicular to the direction of the magnetic field (Tomisaka, Ikeuchi \& Nakamura 1988). A study of a comparison between the density and magnetic field structure observations with theory is planned. \par
Searches of dense cloud cores similar to FeSt 1-457 may reveal that the existence of such physical states is universal. Further modeling of the inclined distorted magnetic fields in FeSt 1-457 may allow us to explore the mass accumulation history of the development of the dense core, which is closely related to the mechanism of core formation in molecular clouds. The formation mechanism for the critical core is an open problem. Quasi-static gravitational contraction or compression by turbulence may be responsible for the formation of the critical core. \par
The present study reveals the 3D magnetic field structure and kinematical stability of the dense starless molecular cloud core FeSt 1-457. The core is found to be in a nearly critical state, which is consistent with the core's starless feature and the kinematic gas motion. FeSt 1-457 is regarded to be in the earliest stage in star formation, i.e., just before the onset of dynamical collapse following the attainment of a supercritical condition. With these well defined physical properties, FeSt 1-457 serves as a good laboratory for the study of the initial conditions of isolated star formation.

\subsection*{Acknowledgement}
We are grateful to the staff of SAAO for their kind help during the observations. We wish to thank Tetsuo Nishino, Chie Nagashima, and Noboru Ebizuka for their support in the development of SIRPOL, its calibration, and its stable operation with the IRSF telescope. The IRSF/SIRPOL project was initiated and supported by Nagoya University, National Astronomical Observatory of Japan, and the University of Tokyo in collaboration with South African Astronomical Observatory under the financial support of Grants-in-Aid for Scientific Research on Priority Area (A) No. 10147207 and No. 10147214, and Grants-in-Aid No. 13573001 and No. 16340061 of the Ministry of Education, Culture, Sports, Science, and Technology of Japan. RK, MT, NK, and KT (Kohji Tomisaka) also acknowledge support by additional Grants-in-Aid Nos. 16077101, 16077204, 16340061, 21740147, 26800111, 16K13791, and 15K05032.


\clearpage 

\begin{figure}[t]  
\begin{center}
 \includegraphics[width=6.5 in]{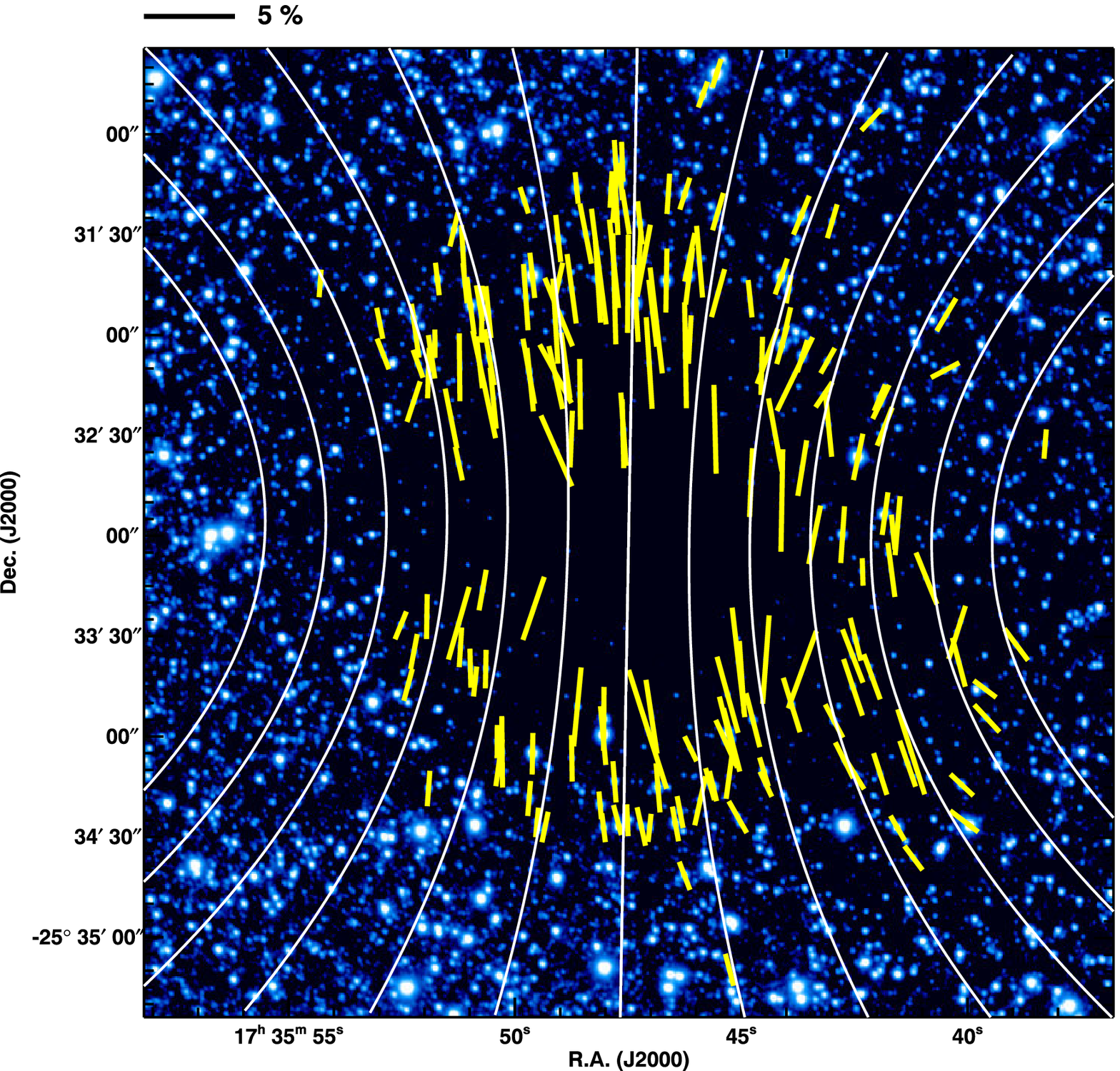}
 \caption{Polarization vectors of FeSt 1-457 after subtraction of the ambient polarization component (yellow vectors). The figure is taken from Paper I. The field of view is the same as the diameter of the core ($288''$). The white lines show the magnetic field direction inferred from fitting with a parabolic function of $y = g + gC{x^2}$, where $g$ specifies the magnetic field lines and $C$ determines the degree of curvature in the parabolic function. The scale of 5$\%$ polarization degree is shown at the top.}
   \label{fig1}
\end{center}
\end{figure}

\clearpage 

\begin{figure}[t]  
\begin{center}
 \includegraphics[width=6.5 in]{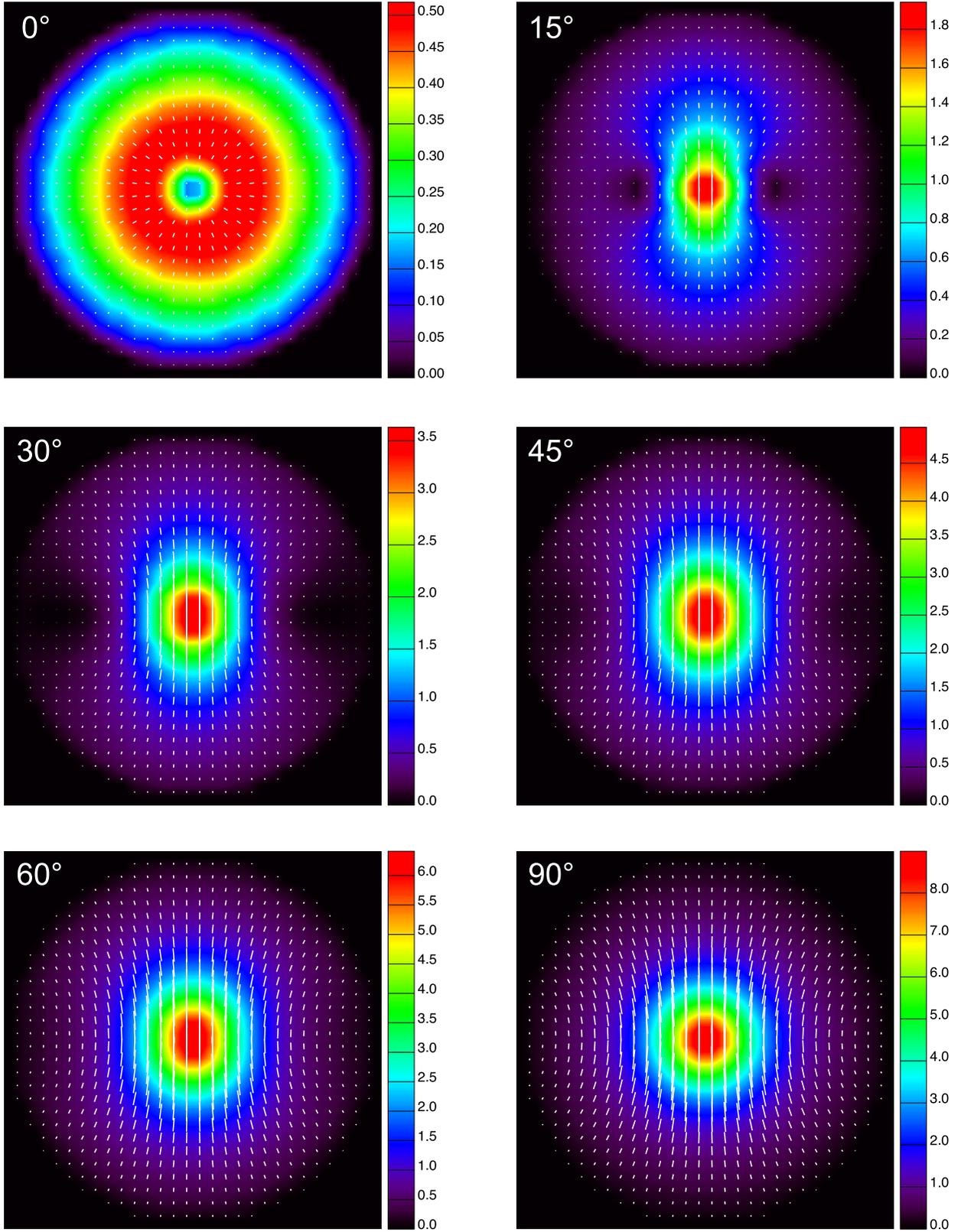}
 \caption{Polarization vector maps of the 3D parabolic model (white vectors). The background color and color bar show the polarization degree. The applied line-of-sight inclination angle is labeled in the upper-left corner of each panel. The 3D magnetic curvature in the model is set to $C=2.5\times10^{-4}$ ${\rm arcsec}^{-2}$ for all the panels.}
   \label{fig1}
\end{center}
\end{figure}

\clearpage 

\begin{figure}[t]  
\begin{center}
 \includegraphics[width=5.0 in]{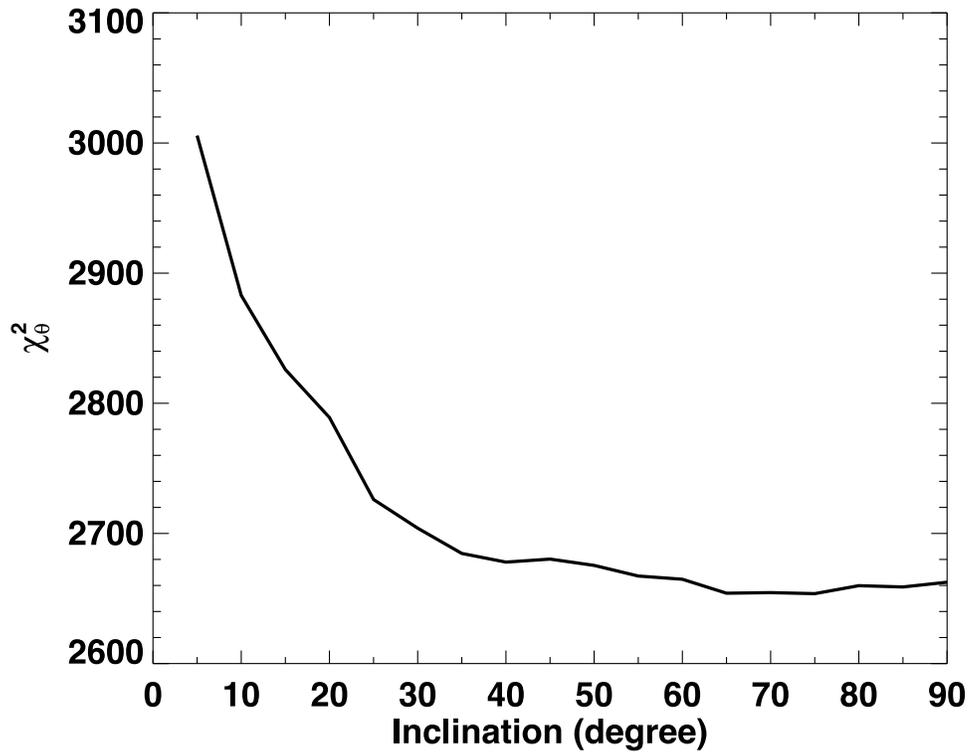}
 \caption{$\chi^{2}$ distribution of the polarization angle $(\chi^2_\theta)$. The best magnetic curvature parameter ($C$) was determined at each $\theta_{\rm inc}$. $\theta_{\rm inc}=0^{\circ}$ and $90^{\circ}$ correspond to the pole-on and edge-on geometry in the magnetic axis.}
   \label{fig1}
\end{center}
\end{figure}

\begin{figure}[t]  
\begin{center}
 \includegraphics[width=5.0 in]{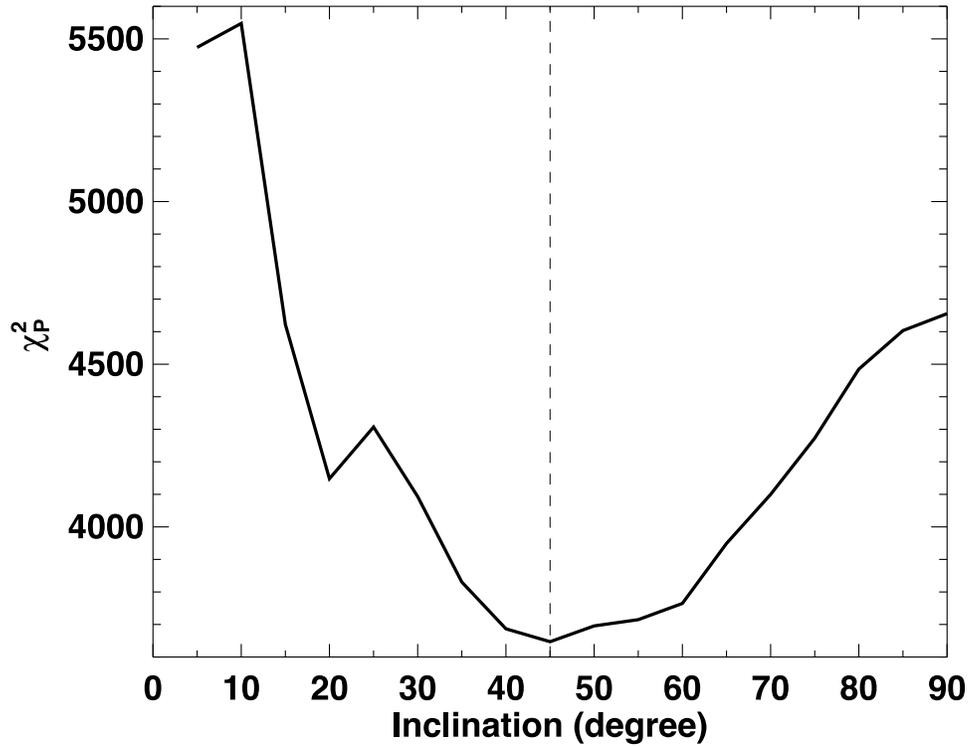}
 \caption{$\chi^{2}$ distribution of the polarization degree $(\chi^2_P)$. $\theta_{\rm inc}=0^{\circ}$ and $90^{\circ}$ correspond to the pole-on and edge-on geometry in the magnetic axis. Calculations of $\chi^2$ in polarization degree were performed after determining the best magnetic curvature parameter ($C$) which minimizes $\chi^2$ in the polarization angle. This calculation was carried out at each $\theta_{\rm inc}$.}
   \label{fig1}
\end{center}
\end{figure}

\clearpage 

\begin{figure}[t]  
\begin{center}
 \includegraphics[width=6.5 in]{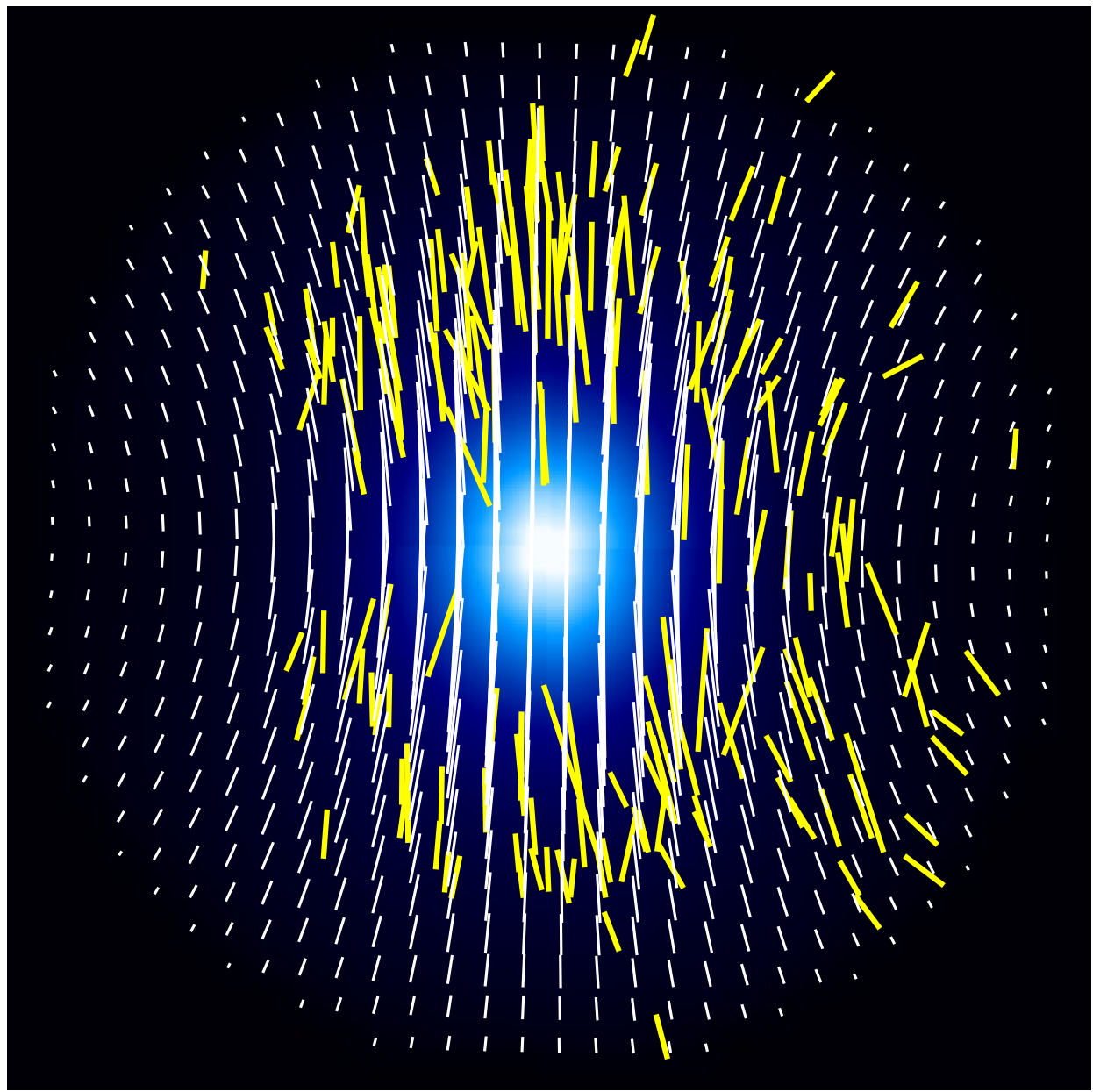}
 \caption{Best-fit 3D parabolic model (white vectors) with observed polarization vectors (yellow vectors). The background color image shows the polarization degree distribution of the best-fit model.}
   \label{fig1}
\end{center}
\end{figure}

\end{document}